# Patent Portfolio Analysis of Cities:

# Statistics and Maps of Technological Inventiveness

Dieter Franz Kogler [a*], Gaston Heimeriks [b], and Loet Leydesdorff [c]

**DRAFT – December 2016**

## Abstract

Cities are engines of the knowledge-based economy, because they are the primary sites of knowledge production activities that subsequently shape the rate and direction of technological change and economic growth. Patents provide a wealth of information to analyse the knowledge specialization at specific places, such as technological details and information on inventors and entities involved, including address information. The technology codes on each patent document indicate the specialization and scope of the underlying technological knowledge of a given invention. In this paper we introduce tools for portfolio analysis in terms of patents that provide insights into the technological specialization of cities. The mapping and analysis of patent portfolios of cities using data of the Unites States Patent and Trademark Office (USPTO) website (at http://www.uspto.gov) and dedicated tools (at http://www.leydesdorff.net/portfolio) can be used to analyse the specialisation patterns of inventive activities among cities. The results allow policy makers and other stakeholders to identify promising areas of further knowledge development and 'smart specialisation' strategies.

[a] *Corresponding author; School of Geography, University College Dublin, Ireland; dieter.kogler@ucd.ie

[b] Innovation Studies, Copernicus Institute of Sustainable Development. Utrecht University, Heidelberglaan 2, 3584 CS Utrecht, The Netherlands; gheimeriks@gmail.com

[c] Amsterdam School of Communication Research (ASCoR), University of Amsterdam, PO Box 15793, 1001 NG Amsterdam, The Netherlands; loet@leydesdorff.net

## 1. Introduction

Cities with their dense mixtures of people and economic activities can be considered the prominent locations of knowledge production and innovation (Bairoch, 1988; Bettencourt *et al.*, 2007; Carlino *et al.*, 2007; Jacobs, 1969). While there has been significant attention for the process of knowledge production in regional or national innovation systems, little consideration has been given to the knowledge produced at specific places. This is a pressing issue because technological knowledge production is highly unevenly distributed over space (Florida 2005), and many cities struggle to replicate the levels of productivity and innovativeness achieved in leading regions. It is difficult for policy-makers to decide how to invest limited resources across the range of leading-edge technologies, especially in cities that are not at the forefront of any specific fields (Heimeriks & Balland 2015).

The present study aims to address the question whether it is possible to empirically specify the unique characteristics of the technological portfolios of cities in terms of technological proximity, distance, and related variety. In pursuit of this objective, we introduce a new instrument for the purpose of mapping and analysing patent portfolios of cities using data available online at the Unites States Patent and Trademark Office website at http://www.uspto.gov and routines at http://www.leydesdorff.net/portfolio. The goal is to analyse the specialisation patterns of inventive activities in different cities.

The starting point of this analysis is the idea that the dynamics of technological knowledge are path- and place-dependent (Heimeriks & Boschma, 2014), and that the current technological portfolio of a region influences the further capacity to develop new technologies (Kogler *et al.*, 2016). From a methodological perspective, the choice for cities as units of analysis is only one among possible applications, but this focus is most relevant from the perspective of innovation studies. Cities have been considered 'innovation machines' (Mellander & Florida, 2016). By choosing cities as units of analysis, we are able to show how theoretical debates about the geography of innovation (Feldman & Kogler, 2010) can be informed using this or other interfaces of patent data for the measurement (e.g., PatentsView at http://www.patentsview.org/web/).

We argue that the barrier between qualitative theorizing and quantitative data mining in the geography of innovation can be overcome by adding statistics to the visualizations of big data (Breschi & Malerba, 2001). On the one hand, the data and statistics enable us to test theoretically informed hypotheses regarding the technological evolution in regional economies (Kogler *et al.*, 2013; Boschma *et al.*, 2014 and 2015), or the diffusion of novel products and processes (Feldman *et al.*, 2015). On the other hand, the analyst is enabled to make one's arguments data-rich and to formulate empirically informed hypotheses.

Bibliographic databases such as the patents at USPTO and elsewhere or databases of scientific publications such as at Google Scholar or the *Science Citation Index*, provide "big" but also "raw" data that may enable us to test hypotheses more effectively than before. However, theoretical notions have to be reformulated with reference to the measurement before one can profit from the potential in this data. These databases are used in bibliometric evaluations; they provide the analyst with two main dimensions (Narin, 1976; Small & Garfield, 1985): (*i*) geographical information in the address field (of authors/inventors or applicants). This address information can be aggregated and reorganized in terms of nations, regions, and cities. The second dimension (*ii*) reflects the intellectual organization of knowledge domains as indicated in groupings of specialized journals, subject categories, keywords or patent classifications. Authors, inventors, and groups of them integrate these two structural dimensions into socio-cognitive actions that one can study in the context of networks of co-authorship or co-invention.

In other words, the data enable us to differentiate between geographical, cognitive, and social maps (Rotolo *et al.*, 2016). Accordingly, concepts of proximity, distance, and related variety can be distinguished in these various dimensions (Frenken *et al.*, 2009). While the geographical dimension can be overlaid onto existing maps (such as Google maps), the intellectual organization is not naturally given so that maps in this dimension have to be carefully constructed. In this study we use the map of aggregated citation relations among 630 Cooperative Patent Classes (CPC) indexed at the USPTO as a baseline for patent portfolio evaluation (Leydesdorff *et al.*, 2014). Subsequently, it will be possible to address the central research question, i.e. how are the patents developed by inventors in specific cities distributed in terms of their technological classes? We measure (*i*) the diversity of portfolios (Rafols & Meyer, 2010; Stirling, 2007; Zhang *et al.*, 2016) and (*ii*) routines are provided to store sets of distributions as vectors in a data matrix that can be used for statistical analysis, for example, in SPSS. Furthermore, (*iii*) input files are generated for the visualization of the portfolios as patent maps using VOSviewer (van Eck & Waltman, 2010).

The following section (Section 2) offers a brief overview of the the relevant literature. The data and methods that will be employed are introduced in Section 3. Section 4 is dedicated to the analysis of results, while the final section will provide a discussion and some concluding suggestions for further research in this line of inquiry, as well as policy recommendations.

## 2. The knowledge production process and spatial patterns of specialization

It has long been recognised that the accumulation of knowledge is central to economic performance (Nelson & Winter, 1982; Romer, 1994; Schumpeter, 1943). In recent years, the importance of knowledge production has further increased due to the process of economic globalisation, the ease of transmitting codified information across geographical space through the Internet, globalisation of corporate R&D, an increase in international collaborations, and the increasing mobility of researchers (Alkemade *et al.*, 2015; David & Foray, 2002; Heimeriks & Vasileiadou, 2008).

Every city has its own, unique knowledge base (Kogler *et al.*, 2013). Cities specialise because existing local skills, infrastructures and institutions facilitate the cumulative and path-dependent character of technological knowledge production (Heimeriks & Boschma, 2014; Martin & Sunley, 2006). The opportunities to diversify into new fields are to a large extent dependent on the existing portfolio of related technological knowledge (Kogler *et al.*, 2016; Boschma *et al.*, 2015). New technologies evolve from the recombination of already existing technological building blocks (Arthur, 2007). Consequently, new technological developments are characterised by a path-dependent process of branching; new technological knowledge is developed from existing knowledge, skills and infrastructures in relation to global developments. From this perspective, the diversity of the technological knowledge base can be considered an important indicator of the innovative potential of a city. Portfolio analysis helps us to understand the technological capabilities that make up a city's patent portfolio.

Cities, and in particular large cities and metropolitan areas, have increasingly been considered as the engines of transition towards a knowledge-based economy (Florida, 2002). Because density in general spurs innovation by bringing people and ideas together and enabling them to combine and recombine in new ways, cities with their dense mixtures of people and economic activities are considered the prominent locations of innovation (Mellander &

Florida, 2016, Camagni, 1999; Hall, 1998). In other words, proximity increases the circulation not only of goods and people, but of ideas as well (Nomaler *et al.*, 2014, Jacobs, 1969). As a consequence, especially metropoles can be expected to benefit from the diversity of human and institutional resources to yield greater output in terms of technological developments (Bettencourt *et al.*, 2007; Glaeser, 2011).

While there has been significant attention for the process of knowledge production in regional or national innovation systems, little consideration has been given to the knowledge produced at specific places. Equally, little is known about how the properties of new knowledge impact upon the performance or future directions of specific firms, sectors, and regions of the economy. Although a number of concepts were introduced that promote local knowledge as a source of regional competitive advantages (Cooke & Leydesdorff, 2006) - including regional innovation systems (Braczyk *et al.*, 1998; Asheim *et al.*, 2012) and the learning region (Morgan, 1997) - only recently theoretical and empirical advances in evolutionary and economic geography have addressed questions regarding the rate and direction of knowledge production, and how this might translate into regional economic wealth (Boschma & Martin, 2010; Kogler, 2015a).

Like other forms of portfolio management (for a recent literature review, see Rafols *et al.*, 2010; Wallace & Rafols, 2015; Zhang *et al.*, 2011), portfolio analysis utilizing patent data can provide insights into the specialization of countries, cities, or knowledge-producing organizations such as universities and firms. A patent prevents an inventor's valuable idea from being commercially implemented by a business rival without penalty. Patents provide legal records of novel, nontrivial, and economic valuable ideas that help drive regional innovation and economic growth. Patents are essential for avoiding market failure that is likely to occur in the absence of intellectual property rights due to the positive externalities generated by novel products and processes, and knowledge in general; in essence they can be considered as vital instruments in the quest for technological development (Greenhalgh & Rogers, 2010).

As noted, we introduce an instrument for the purpose of mapping and analysing patent portfolios of cities. The theoretical objective is to understand the specialisation patterns of inventive activities at the city level. The longer-term perspective is the idea that the dynamics of technological knowledge are path and place dependent (Heimeriks & Boschma, 2014;

Martin & Sunley, 2006), and that the current technological portfolio of a city or region influences its further capacity to produce new technologies (Kogler *et al.*, 2016).

Following from this brief discussion of the relevant literature, we expect that cities can be characterised by distinct technological portfolios. From a policy perspective, portfolio analyses inform policy makers in their mission to make best use of the existing technological strengths of cities. To the best of our knowledge, the results will for the first time provide the opportunity for the comparison of city's individual knowledge spaces along various measures and dimensions. Furthermore, the suggested approach offers measures of technological distance that should further the understanding of the adjacent possibilities, i.e. the prospect of developing new capabilities in unoccupied knowledge domains that are adjacent to existing ones in the local knowledge space (Kogler *et al.*, 2013).

## 3. Data and Methods

### *3.1. Data*

Patent data provide a wealth of information pertaining to the creation and diffusion of technical knowledge in cities, regions, and countries (Usai, 2011). Patents can be used for analyzing patterns of invention along the dimensions of locations, technology classes, and organizations. However, the disadvantages of patents as overall measures of economic and inventive activity are well known (Scherer, 1984; Pavitt, 1985; Griliches, 1990; Archibugi & Pianta 1996; OECD 2009). One of these refers to the stark variation in the propensity to patent among economic sectors; patenting is prevalent in what are considered high-tech or knowledge-intensive industries, e.g. information and communication technologies, chemicals, pharmaceuticals, and measuring and optical instruments (Kogler, 2015b). Another limitation pertains to the skewed distribution of the value of patents (Zeebroeck 2011; Zeebroeck & Van Pottelsberghe de la Potterie, 2011).

Notwithstanding these limitations, patents can provide important insights into the individuals and organizations actively engaged in inventive activity in technologies where the protection of intellectual property is a key aspect (Levin *et al.*, 1987). Patent databases are widely available online (Kim & Lee, 2015); for the present analysis the freely accessible interface of the United States Patent and Trade Office (USPTO) was utilized in order to download sets of patents in batch jobs on the basis of composed search strings. Among the various databases,

USPTO data can be considered the most appropriate reflection of technological inventiveness across jurisdictions, and therefore this data has been widely applied in cross-country studies (Fu and Yang, 2009; Johansson *et al.*, 2015).

We make use of the Cooperative Patent Classification (CPC) system. CPC is based on an agreement between USPTO and the European Patent Office (EPO) about the indexing of patents. However, CPC are identical in the first four digits to the older IPC (that is, International Patent Classes).[1] Our routines provide four and three-digit maps, but the analysis is pursued at the four-digit level. At the four-digit level the IPC classification system contains 630 distinct technology categories, and the map is based on citation patterns among the USPTO patents grouped according to the IPC classes they are assigned to (cf. Bowen & Liu, in press).

Given the explorative nature of this research, four cities in each of five different countries were selected as examples. The objective behind this specific sample of cities is to cover sufficient variety in different dimensions. France, for example, is a larger country within the EU with a centralized structure where Paris is the primary metropolitan area. The Netherlands on the other hand is a smaller member state where the urban hierarchy is not as pronounced. In the mix are also cities located in China, Israel, and the U.S. The five countries and selected cities are listed in Table 1.

**Table 1**: Twenty cities in five countries

| **Country** | **Cities** |
|---|---|
| China | Beijing, Shanghai, Nanjing, Dalian |
| France | Paris, Marseille, Grenoble, Toulouse |
| Israel | Jerusalem, Tel Aviv, Haifa, Beer Sheva |
| Netherlands | Amsterdam, Rotterdam, Eindhoven, Wageningen |
| USA | Boston, Atlanta, Berkeley, Boulder |

Patents issued in 2014 were downloaded, since at the time of the retrieval (October 2015), the year 2015 was not yet complete. We use the database of granted patents (at http://patft.uspto.gov/netahtml/PTO/search-adv.htm) because this data is of higher quality than patent applications (at http://appft.uspto.gov/netahtml/PTO/search-adv.html). The application-grant lag distribution for USPTO patents shows that most patents are granted within 3 years of their application (Hall *et al.*, 2001), but significant outliers remain (Popp *et al.*, 2003).

[1] IPC was replaced with the Cooperative Patent Classification by USPTO and the European Patent Organization (EPO) on January 1, 2013. CPC contains new categories classified under "Y" that span different sections of the IPC in order to indicate new technological developments (Scheu *et al.*, 2006; Veefkind *et al.*, 2012).

The search string is like "ic/amsterdam and icn/nl and isd/2014$$" for non-American cities or "ic/boston and is/ma and isd/2014$$" using the state abbreviation instead of the country name for cities in the U.S.A. The retrieval is listed in Table 2. Note that we did not limit the application dates backward.

**Table 2**: Retrieval rates for four cities in five countries.

| ***China*** | | ***France*** | | ***Israel*** | | ***Netherlands*** | | ***USA*** | |
|---|---|---|---|---|---|---|---|---|---|
| Beijing | 2,122 | Paris | 1,336 | Jerusalem | 283 | Amsterdam | 253 | Boston | 874 |
| Shanghai | 1,669 | Marseille | 13 | TelAviv | 876 | Rotterdam | 102 | Atlanta | 1,166 |
| Nanjing | 192 | Grenoble | 422 | Haifa | 776 | Eindhoven | 884 | Berkeley | 854 |
| Dalian | 39 | Toulouse | 324 | BeerSheva* | 55 | Wageningen | 43 | Boulder | 910 |

* The search string for BeerSheva is: "(ic/beer-sheva or ic/beersheva) and icn/il and isd/2014$$"

The level of precision obtained from searching with city names is not controlled. Some cities are administratively underbounded (e.g., Amsterdam, Rotterdam) and may have suburbs that are not captured by the search while contributing to the metropolitan labour market, whereas other cities are overbounded (e.g., Boulder, CO). In the USA, Core Based Statistical Areas (CBSA) are defined by the US Office of Management and Budget (OMB). A CBSA is a group of adjacent areas that are socioeconomically close to an urban center. However, series of attempts at constructing a European counterpart to the metropolitan region concept of the US are still short of results, which could be used for the purpose of comparing the scientific base of large cities (Grossetti *et al.*, 2014; Maisonobe *et al.*, 2016).

The composition of CBSA in terms of counties can be found at http://www.uspto.gov/web/offices/ac/ido/oeip/taf/cls_cbsa/cbsa_countyassoc.htm. For the four cities in the USA listed in Table 1, we additionally explore the effect of this alternative definition and elaborate the analysis for the metropolitan definition of Boston. The complete search string for the CBSA "Boston-Cambridge-Quincy, MA-NH," for example, is "(ic/(Essex OR Middlesex OR Norfolk OR Plymouth OR Suffolk OR Boston OR Cambridge) AND IS/MA) OR (ic/(Quincy OR Rockingham OR Strafford) AND IS/NH) AND ISD/2014$$", for example, leads to a retrieval of 2,265 records as against 874 patents for the original search with only "Boston" (MA) as city name (Table 3).

**Table 3**: Search strings and retrieval for four metropolitan regions in the USA.

| *City* | *Search string for the Metropolitan Area (CBSA)* | *Retrieval* |
|---|---|---|
| Boston | (ic/(Essex OR Middlesex OR Norfolk OR Plymouth OR Suffolk OR Boston OR Cambridge) AND IS/MA) OR (ic/(Quincy OR Rockingham OR Strafford) AND IS/NH) AND ISD/2014$$ | 2,265 |
| Atlanta | IS/GA and isd/2014$$ and ic/(Atlanta OR "Sandy Springs" OR Marietta OR Barrow OR Bartow OR Butts OR Carroll OR Cherokee OR Clayton OR Cobb OR Coweta OR Dawson OR DeKalb OR Douglas OR Fayette OR Forsyth OR Fulton OR Gwinnett OR Haralson OR Heard OR Henry OR Jasper OR Lamar OR Meriwether OR Newton OR Paulding OR Pickens OR Pike OR Rockdale OR Spalding OR Walton) | 1,526 |
| Berkeley | IS/CA and isd/2014$$ and ic/("San Francisco" OR Oakland OR Fremont OR Alameda OR "Contra Costa" OR Marin OR "San Mateo")* | 10,207 |
| Boulder | IS/CO and isd/2014$$ and ic/Boulder | 910 |

* Addition of "OR Berkeley" augments the retrieval with 534 patents to 10,741. Berkeley is part of the county Alameda in the CBSA of San Francisco-Oakland-Fremont, CA.

### *3.2. Methods*

Dedicated routines were written which enable the user to download retrieved sets in batches of 1,000 patents. The routines generate files for the mapping as an overlay using VOSviewer for the visualization, and files for network analysis and visualization using Pajek. The various fields in the USPTO records are parsed and then organized in a series of databases that can be related using, for example, MS Access. The procedure is further specified in Appendix I and the routines are available online at http://www.leydesdorff.net/software/patents.

If not yet initially present the routine generates additionally the files matrix.dbf and rao.dbf, which are incrementally extended with rows and columns in each subsequent run. After each retrieval, a column variable is added to the file matrix.dbf containing the distribution of the 630 CPC/IPC classes in the additional document set under study. This matrix can be read into Excel (or SPSS, etc.) for statistical analysis. Analogously, a row variable is added after each run to the file rao.dbf containing diversity measures (see below) as variables. As noted, these files are generated *de novo* if previously absent.

The additional routine ipc2cos.exe reads the file matrix.dbf and produces co-occurrence matrices which can be used for further analysis in programs such as Pajek or UCInet.

Normalization using the cosine values (in "cosine.net") brings the latent structure to the foreground,[2] whereas visualizations based on the non-normalized file ("coocc.dat") tend to show the relational variation.

### *3.3. Diversity; "related variety"*

We are not only interested in the size of the patent portfolio of cities, but also in the diversity contained within the portfolio. Diversity may refer to both the number of different categories (e.g., technology classes) and the disparity among these categories. Rao-Stirling diversity is a measure that takes into account both the variety and the disparity in a patent portfolio under study across the IPC classes. In other words, the variety is considered as ecologically related in terms of the categories (Frenken *et al.*, 2007).

The resulting Rao-Stirling diversity is defined as follows (Rao, 1982; Stirling, 2007):

$$\Delta = \sum_{ij} p_i p_j d_{ij} \qquad (1)$$

where $d_{ij}$ is a distance or disparity measure between two categories *i* and *j*—the categories are in this case IPC classes—and $p_i$ is the proportion of elements assigned to each class *i*. As the disparity measure, we use (1 – *cosine*) since the cosine values among all aggregated IPC is used for constructing the base map of three and four digits. Jaffe (1986, at p. 986) proposed the cosine between the vectors of classifications as a measure of "technological proximity".

Zhang *et al.* (2016) argues that ${}^2D^S$ provides a true diversity measure that outperforms Rao-Stirling diversity (Δ) because ${}^2D^S = 2.0$ is twice as diverse as ${}^2D^S = 1.0$. In their Equation 6 (at p. 1260), however, these authors formulate:

$${}^2D^S = 1/(1 - \Delta) \qquad (2)$$

where $\Delta$ is the Rao-Stirling diversity. In other words, the transformation is monotonic and the value of ${}^2D^S$ follows directly from that of the Rao-Stirling diversity using Eq. 2. This improved measure varies from 1 to ∞ when $\Delta$ varies from 0 to 1. Both measures are provided for each case in the file "rao.dbf". Note that these are diversity measures of each portfolio in terms of the composition of IPC classes at the four digit level.

[2] The cosine is similar to the Pearson correlation except that the distributions are not *z*-normalized to the mean. Since the patent distributions are non-normal (but skewed), this measure is more appropriate (Ahlgren *et al.*, 2003).

## 4. Results

### *4.1. Comparisons among individual cities*

The routines outlined in the previous section enable a comparison of sets (in our case, cities) within clusters (for example, countries) or across cities and countries using multivariate analysis of the matrix, which is incrementally constructed during subsequent runs. This statistical analysis will be the subject of the next section; but let us focus now on an example that shows how a more qualitative approach using visualizations informs the analysis. Note that visualization is not an analytical technique. However, it allows one to recognize patterns which can then further be tested. In other words, visualizations serve the generation of hypotheses more than statistics.

Zooming in on two French cities, Figure 1 shows a comparison between Paris (Figure 1A at the top) and Toulouse (Figure 1B at the bottom) overlaid on the global map of IPC. As noted (in Table 2 above), 1,336 USPTO patents were granted in 2014 to inventors with a Paris address, whereas this number was 324 for Toulouse. However, one should keep in mind that Paris is under-bound as the center of Île-de-France, a larger metropolitan area made up of nine administrative departments (Paris, Essonne, Hauts-de-Seine, Seine-Saint-Denis, Seine-et-Marne, Val-de-Marne, Val-d'Oise, and Yvelines).[3] Figure 1A, however, shows the typical pattern of patenting in a large-scale metropolitan region across the map: 226 of the 630 classes are populated. For Toulouse, a significantly smaller urban center the number of classes occupied is just 110.

[3] Because of the diacritical characters searching with these names is difficult in USPTO; but we found one patent with "Essone" in the address field, three with "Val-de-Marne", and seven with "Yvelines," granted in 2014.

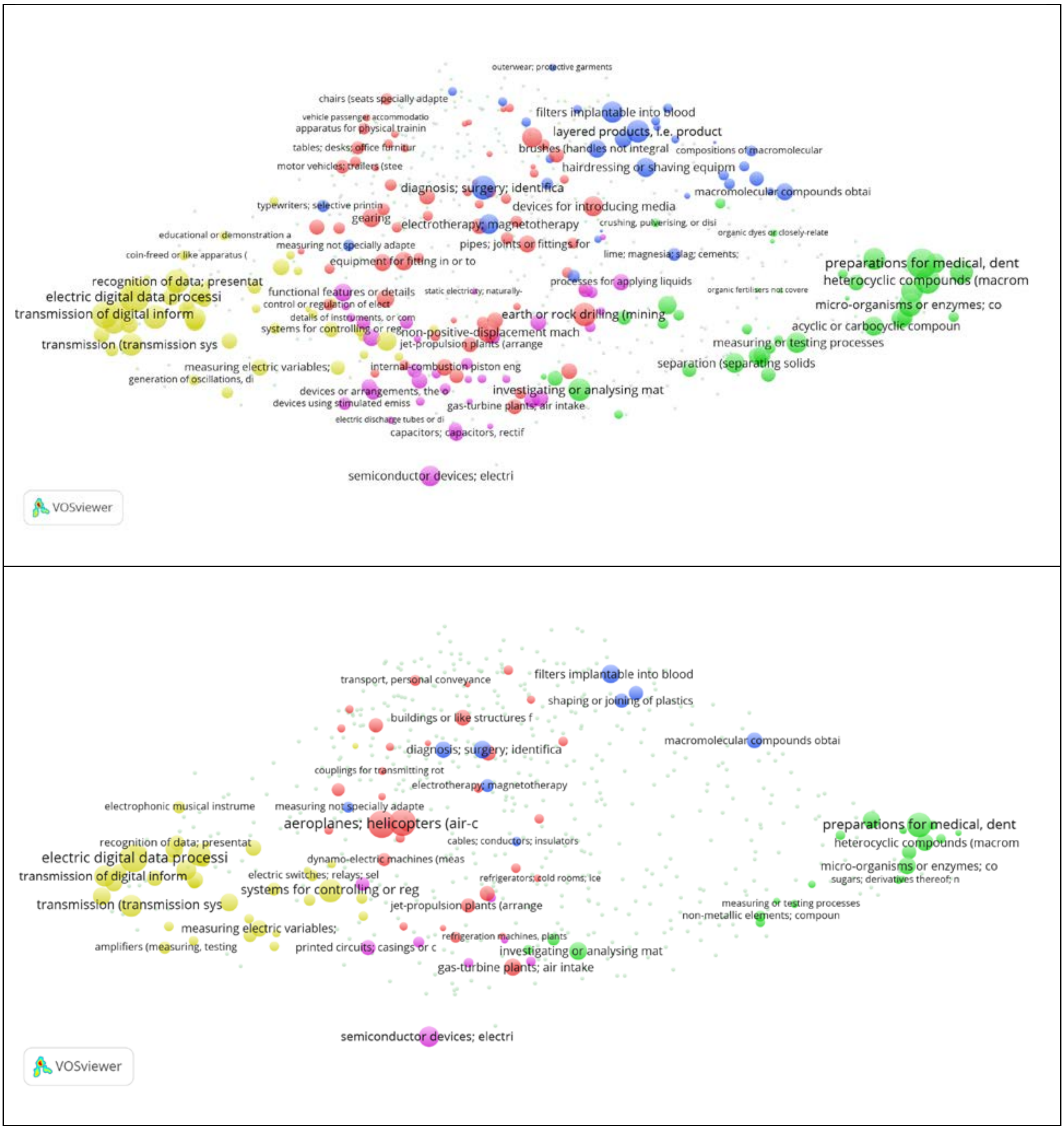

**Figure 1**: Overlays of patent portfolios for Paris (France) and Toulouse (France) in 2014. The map for Paris can be web-started at http://www.vosviewer.com/vosviewer.php?map=http://www.leydesdorff.net/software/patents/paris.txt&label_size_variation=0.3&scale=1.1; the one for Toulouse at http://www.vosviewer.com/vosviewer.php?map=http://www.leydesdorff.net/software/patents/toulouse.txt&label_size_variation=0.3&scale=1.1.

In both figures, a cluster of bio-medical patents can be found on the right side. This cluster is found in almost all western cities and regions (Leydesdorff *et al.*, 2016a). In Toulouse, however, this cluster is disconnected from the largest component of 86 patent classes representing various forms of engineering and related techno-sciences. Figure 2 shows the network visualization of this component (extracted from the set). In this local representation—no longer projected onto the global map of 630 categories—the airplane industry, which is of significant size in Toulouse due to the presence of Airbus, is visible in a cluster of patents at the bottom-right, but is somewhat distanced from the other technology clusters in the city.

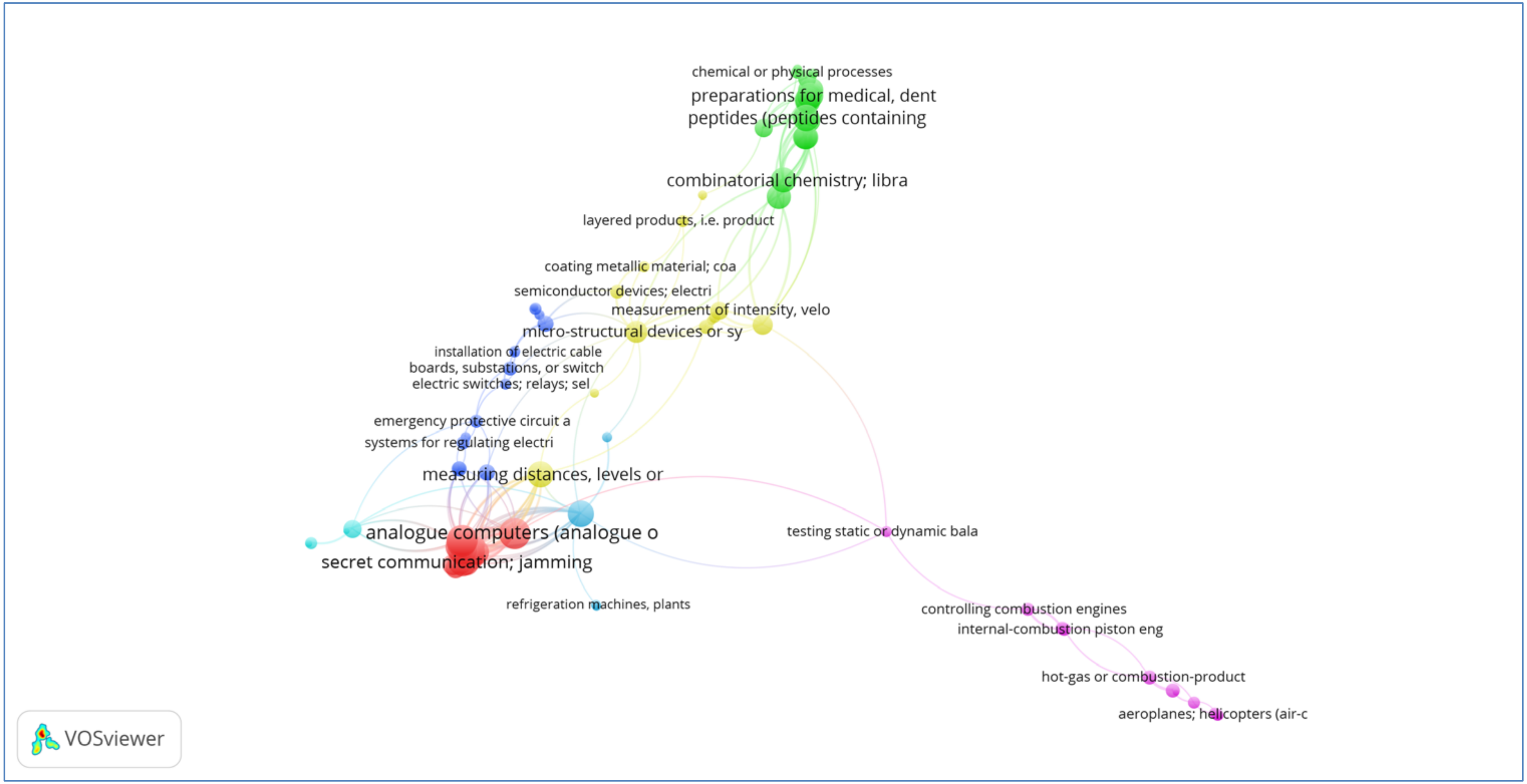

**Figure 2**: Largest component of 86 (among 110) patent classes with an inventor address in Toulouse, France. The map can be web-started at http://www.vosviewer.com/vosviewer.php?map=http://www.leydesdorff.net/software/patents/toul86.txt&network=http://www.leydesdorff.net/software/patents/toul86n.txt&label_size_variation=0.3&scale=1.25&colored_lines&curved_lines&n_lines=10000 ; threshold: cosine > 0.2.

In Table 4, the patent portfolios of Paris and Toulouse are compared as local networks of co-occurring classifications.[4] Values for Toulouse are lower than for Paris in most cases; but centralization is higher for Toulouse than Paris. The density of the network of Toulouse is almost twice as high when compared with Paris. In other words, the clustering of technological knowledge as indicated by the classification codes found in patents generated by inventors residing in Toulouse is more concentrated when compared with Paris where patents are more diverse and spread across the network.

**Table 4**: Network cohesion measures of the portfolios of Paris and Toulouse (using UCInet).

| ***UCInet*** | ***Network Cohesion Measures*** | ***Paris*** | ***Toulouse*** |
|---|---|---|---|
| **1** | Avg Degree | 8.159 | 6.855 |
| **2** | Indeg H-Index | 24 | 16 |
| **3** | Deg Centralization | 0.138 | 0.17 |
| **4** | Out-Central | 0.138 | 0.168 |
| **5** | In-Central | 0.138 | 0.168 |
| **6** | Density | 0.036 | 0.063 |
| **7** | Components | 25 | 16 |
| **8** | Component Ratio | 0.107 | 0.138 |
| **9** | Connectedness | 0.744 | 0.613 |
| **10** | Fragmentation | 0.256 | 0.387 |
| **11** | Closure | 0.666 | 0.748 |
| **12** | Avg Distance | 4.184 | 4.032 |
| **13** | SD Distance | 1.761 | 2.098 |
| **14** | Diameter | 12 | 11 |
| **15** | Breadth | 0.775 | 0.786 |
| **16** | Compactness | 0.225 | 0.214 |

In summary, the analysis of the networks in Paris and Toulouse shows a clear distinction between a large metropole with a diverse technological knowledge base and a more specialised medium sized city. From a policy perspective, this raises the issue what strategies are available for these two locations? For Toulouse, an obvious strategy seems to be to identify options for related diversification. Given its strong pattern of specialisation, adjacent technological opportunities can be identified. For Paris, its advantage lies in the diversity of its technological knowledge base. In addition to expanding its many technological strengths through related diversification, Paris seems well positioned to further develop a comparative advantage in complex technological knowledge that requires a recombination of diverse technological building blocks at both the global and local levels (Fleming & Sorenson, 2001).

[4] UCInet enables the user to generate these network statistics in a single pass.

*4.2. Comparisons at the level of the set*

In addition to network analysis of co-classifications, the routine enables us to compare portfolios by considering patent classes as attributes to the cities as units of analysis. To this end a matrix is incrementally constructed: in each run an additional variable is added. This variable has values larger than zero for the classes which are attributed. For example, in the case of Paris 226 classes are used, and (630 – 226 =) 404 classes are empty. Unlike in the previous analysis where the focus has been on the relations among patent classes, this matrix enables one to analyse *correlations* between portfolios of cities. In contrast to the relational (network) analysis, correlations span a vector-space in which one can distinguish densities as principal components.

Using (5 * 4 =) 20 cities, the result is a matrix of 630 IPC classes versus 20 cities. This matrix can be used as input for multi-variate analysis in a statistics program such as SPSS. The portfolios of Paris and Toulouse are correlated with Pearson $r = .691$ ($p < .01$); the Spearman rank-order correlation $\rho = .472$ ($p<.01$). The lower value of the rank-order correlation indicates that the portfolios have different foci; but overall there is a lot of correspondence. However, note that the correlation is partially caused by the large number of zeros. The rank-order correlation for the 83 classes attributed to both cities is .620 ($p<.01$); the cosine—a non-parametric equivalent of the Pearson correlation (Ahlgren *et al.*, 2003)—is 0.703.

**Table 5**: Discriminant analysis of 20 cities in terms of 630 patent classes.

**Classification Results**[a]

| country | | | Predicted Group Membership | | | | | Total |
|---|---|---|---|---|---|---|---|---|
| | | | France | Israel | China | USA | Netherlands | |
| Original | Count | France | 3 | 0 | 1 | 0 | 0 | 4 |
| | | Israel | 0 | 3 | 1 | 0 | 0 | 4 |
| | | China | 0 | 0 | 4 | 0 | 0 | 4 |
| | | USA | 0 | 0 | 0 | 4 | 0 | 4 |
| | | Netherlands | 0 | 0 | 1 | 0 | 3 | 4 |
| | % | France | 75.0 | 0.0 | 25.0 | 0.0 | 0.0 | 100.0 |
| | | Israel | 0.0 | 75.0 | 25.0 | 0.0 | 0.0 | 100.0 |
| | | China | 0.0 | 0.0 | 100.0 | 0.0 | 0.0 | 100.0 |
| | | USA | 0.0 | 0.0 | 0.0 | 100.0 | 0.0 | 100.0 |
| | | Netherlands | 0.0 | 0.0 | 25.0 | 0.0 | 75.0 | 100.0 |

a. 85.0% of original grouped cases correctly classified.

Table 5 shows the result of discriminant analysis using the portfolios of cities as predictors of the national origins. Since the latter is known *ex ante*, one can note that the statistical prediction is perfect (100%) for the USA and China. France, Israel and the Netherlands each contain one city with a profile that is sorted by the routine into the Chinese group. These are respectively: Marseille, Beersheva, and Wageningen. Consequently, the discrimination is not statistically significant; the Dutch cities, notably, entertain portfolios which are close to the ones of China (Figure 3). Nevertheless, a national character of the portfolios is weakly indicated. The USA is the outlier in Figure 3, but this may find its origin in the utilization of USPTO data.

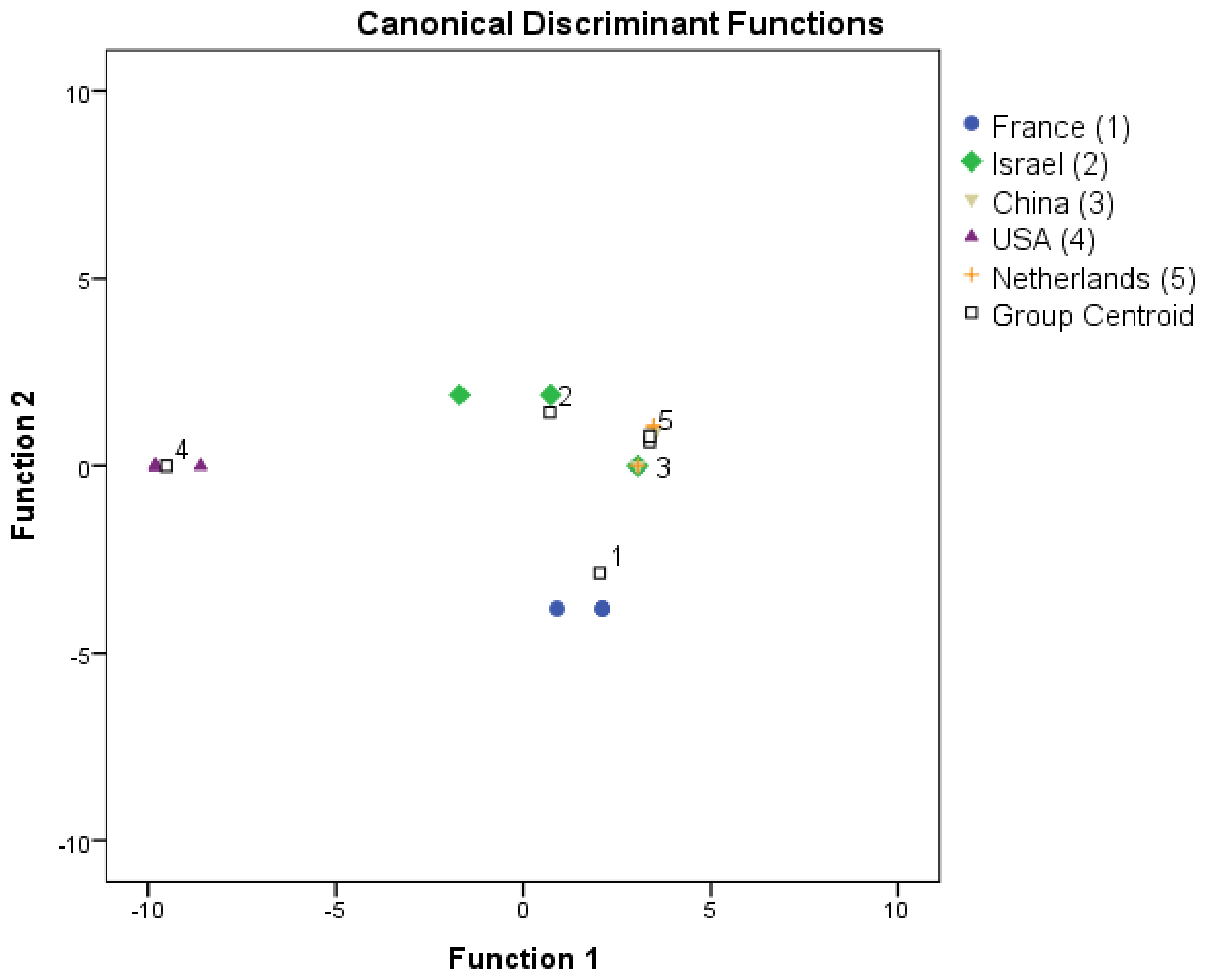

**Figure 3**: All-groups scatterplot of the twenty cities in four countries using canonical discriminant functions.

Table 6 shows the results of factor analysis of the 20 cities as variables.[5] Three factors explain 78.3% of the variance. Factor 1 assembles the cities with a portfolio focusing on engineering; factor 2 indicates a prevailing portfolio in the bio-medical domain, whereas only Dalian (China) and Amsterdam (the Netherlands) score highest on factor 3. This factor is more difficult to designate.

**Table 6**: Varimax-rotated factor matrix of the patent portfolios of 20 cities.

| | Component | | |
|---|---|---|---|
| | 1 | 2 | 3 |
| Beijing | **.923** | .174 | .107 |
| Haifa | **.902** | .291 | |
| Berkeley | **.884** | .340 | |
| Tel Aviv | **.876** | .377 | |
| Atlanta | **.858** | .369 | |
| Boulder | **.818** | .382 | |
| Shanghai | **.802** | .358 | .218 |
| Nanjing | **.737** | .370 | .271 |
| Grenoble | **.729** | | .135 |
| Toulouse | **.662** | .454 | |
| Eindhoven | **.613** | | |
| Marseille | **.521** | **.500** | .170 |
| Wageningen | | **.816** | |
| Boston | .456 | **.811** | .176 |
| Paris | .471 | **.810** | .150 |
| Rotterdam | .223 | **.799** | |
| Beersheva | .451 | **.779** | |
| Jerusalem | .611 | **.731** | |
| Dalian | | | **.951** |
| Amsterdam | .244 | **.622** | **.634** |

Extraction Method: Principal Component Analysis. Rotation Method: Varimax with Kaiser Normalization. a. Rotation converged in 5 iterations.

While Beijing has the highest loading on factor 1, Dalian has a very different pattern of patenting. In order to further understand the difference between these two cities, one could, for example, map Dalian versus Beijing analogously as we mapped Toulouse versus Paris in

[5] We use the transposed matrix because factor scores are more difficult to read, while factor scores do not vary between -1 and +1.

Figure 1. The factor analysis thus suggests a way forward if one is particularly interested in Chinese portfolios, or in evaluating differences amongst places altogether.

### *4.3. A map of the 20 cities*

The matrix of 20 cities versus 630 patent classes enables us also to make a distance matrix using for example the cosine values between the vectors. The *cosine* is a similarity measure, but (1 – *cosine*) provides us with a dissimilarity measure or distance. Feeding these distances into a visualization program one can map and cluster the cities. In other words, these distributions are normalized. By adding geo-codes to the cities, one would also able to map the cities geographically (Leydesdorff & Bornmann, 2012).

.

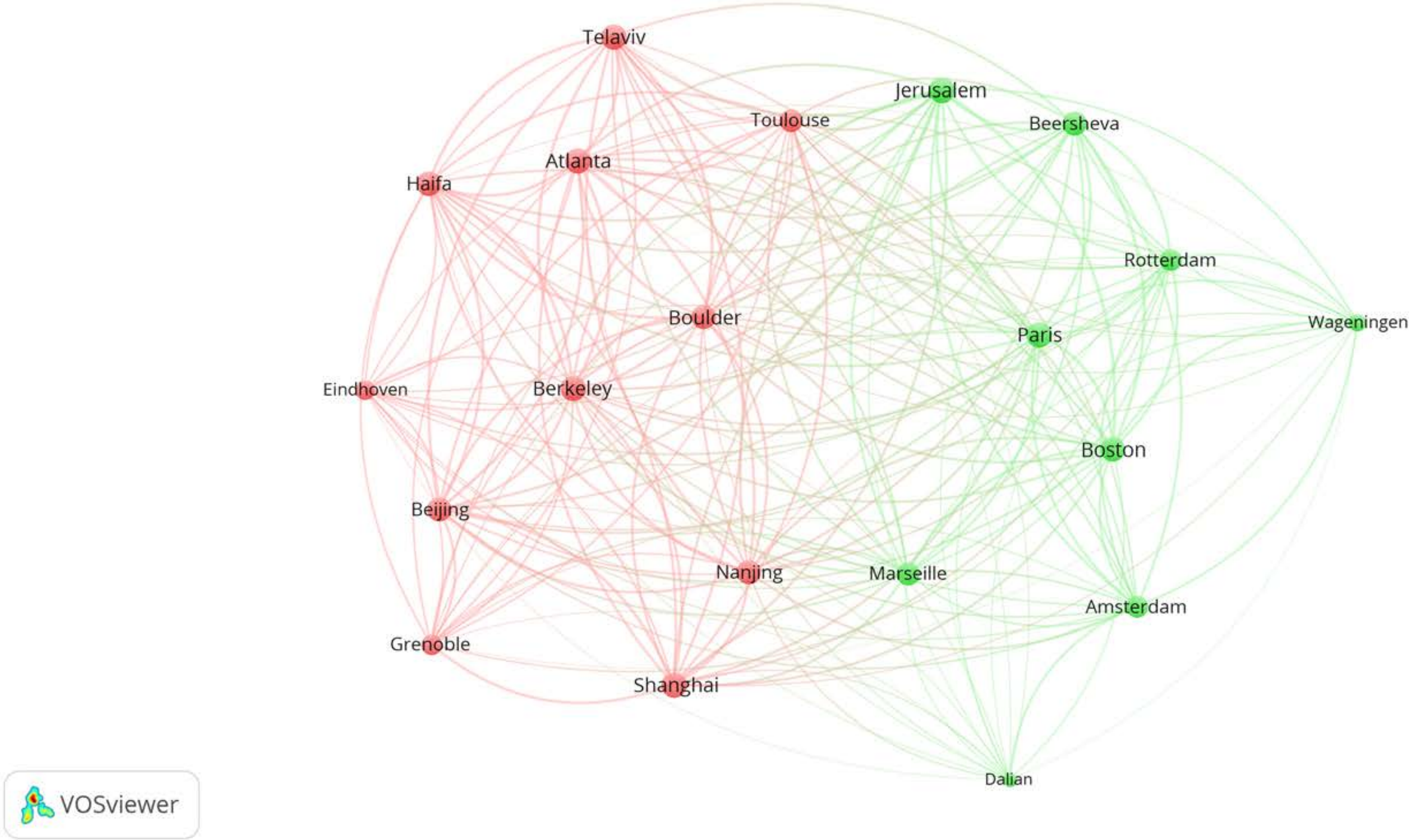

**Figure 4**: Cosine-normalized network among 20 cities; VOSviewer is used for the mapping and clustering. The map can be web-started at http://www.vosviewer.com/vosviewer.php?map=http://www.leydesdorff.net/software/patents/cos_map.txt&network=http://www.leydesdorff.net/software/patents/cos_net.txt&label_size_variation=0.3&scale=1.40&colored_lines&curved_lines&n_lines=10000&line_size_variation=0.55

Using VOSviewer for the clustering and the mapping, two types of portfolios are distinguished, as indicated with green and red in Figure 4. The divide can be characterized as American-Pacific versus American-Atlantic portfolios. Leydesdorff *et al*. (2016a) found a similar divide when analysing university patents at the level of countries. An alternative characterization, however, in terms of engineering versus bio-medicine explains also why Toulouse, Grenoble, and Eindhoven are part of the red-coloured cluster. The factor-analysis (Table 6) informs us that these cities are weakly loading on the relevant factor 1. In this two-cluster solution, Dalian sides in the vicinity of Amsterdam and Marseille in the Atlantic cluster. Note that these two European cities showed interfactorial complexity.

### *4.4. Related Variety*

In Table 6, we rank the 20 cities in terms of decreasing Rao-Stirling diversity, and compare this with the portfolio analysis of these 20 cities using scientific publications in the Web of Science provided in a previous study (Leydesdorff *et al*., 2016b). As explained in the methods section, Rao-Stirling diversity can be considered as a measure of "related variety" (Castaldi *et al*., 2015; Frenken *et al*., 2007). The measure is also called "quadratic entropy" or "ecological entropy" (Izsák & Papp, 1995; Rao, 1982; Ricotta & Szeidl, 2006). The ecological distance ($d_{ij}$) between species *i* and *j* is multiplied by their variety ($p_i$ * $p_j$).[6] Variety which is "related"—such as in an ecological niche—is thus accounted for differently from variety which is "unrelated." Unlike Castaldi *et al.* (2015), this formulation does not require the definition of given categories, such as nested levels of the IPC, but only a distance measure such as (1 – *cosine*) (Jaffe, 1989).

We use (1 – $cosine_{ij}$) as a measure of dissimilarity or distance in this case; the cosine is provided between each two of the 630 IPC4 classes in a file at http://www.leydesdorff.net/ipcmaps/cos_ipc4.dbf. In Table 7, the resulting values are listed in rank order. In the right half of the table, the values of Δ are provided from a previous study in which portfolios of journals were analysed for the same 20 cities (Leydesdorff *et al*., 2016b, Table 3, p. 746).

[6] ($p_i$ * $p_j$) is the Gini-Simpson index. The Gini-Simpson is equal to the complement to one of the Herfindahl–Hirsch index or equivalently the Simpson index (Stirling, 2007).

**Table 7**: Rao-Stirling diversity for 20 cities in USPTO and WoS, respectively.

| **USPTO (a)** | **Rao Δ (b)** | **N (c)** | **WoS (d)** | **Rao Δ (e)** | **N (f)** |
|---|---|---|---|---|---|
| Shanghai | 0.8894 | 1,669 | Haifa | 0.3277 | 3,408 |
| Eindhoven | 0.8725 | 884 | Beer Sheva | 0.3138 | 1,905 |
| Paris | 0.8702 | 1,336 | Tel Aviv | 0.3128 | 4,206 |
| Rotterdam | 0.8684 | 102 | Paris | 0.3112 | 24,877 |
| Dalian | 0.8653 | 39 | Marseille | 0.3081 | 5,293 |
| Boulder | 0.8637 | 910 | Toulouse | 0.3043 | 5,899 |
| Toulouse | 0.8630 | 324 | Jerusalem | 0.2981 | 3,414 |
| Amsterdam | 0.8557 | 253 | Shanghai | 0.2915 | 29,166 |
| Nanjing | 0.8532 | 192 | Atlanta | 0.2846 | 14,296 |
| Grenoble | 0.8510 | 422 | Eindhoven | 0.2838 | 2,554 |
| Beer Sheva | 0.8458 | 55 | Amsterdam | 0.2737 | 13,451 |
| Boston | 0.8447 | 874 | Berkeley | 0.2719 | 8,868 |
| Atlanta | 0.8446 | 1,166 | Beijing | 0.2621 | 58,032 |
| Berkeley | 0.8215 | 854 | Nanjing | 0.2547 | 17,713 |
| Jerusalem | 0.8116 | 283 | Grenoble | 0.2457 | 5,564 |
| Beijing | 0.8047 | 2,122 | Boulder | 0.2216 | 5,274 |
| Tel Aviv | 0.7748 | 876 | Boston | 0.2091 | 31,182 |
| Haifa | 0.7578 | 776 | Wageningen | 0.2010 | 3,178 |
| Marseille | 0.7061 | 13 | Dalian | 0.2004 | 5,023 |
| Wageningen | 0.5426 | 43 | Rotterdam | 0.1932 | 5,721 |

The numbers of patents and publications (*N* in Table 6) are significantly correlated ($r = 0.753$; $p < 0.01$). However, this correlation may be spurious: both numbers can be expected to co-vary with size. The diversity, however, is negatively correlated ($r = -0.102$; *n.s.*). In other words, patenting and publishing operate in two different selection environments.[7]

For example, the Israeli cities Haifa, Beer-Sheva, and Tel Aviv were ranked as the highest on diversity in terms of journal publications (in WoS), but Haifa and Tel Aviv are among the lowest in terms of diversity among the patents. In other words, these cities contain knowledge-producing institutions (e.g., universities) which are prolific and publish in a large number of fields. However, their patenting portfolios are specific. The selection mechanisms for patents are very different from those for publications.

[7] We use Zhang *et al.*'s (2016) diversity measure ($^2D^S$) for estimating this correlation since $^2D^S = {}^1/_{1-\Delta}$ measures "true diversity" with which one is allowed to calculate as a variable at the ratio scale.

### *4.5. Cities and Metropolitan Areas*

In the following, we added the retrieval for the CBSA 14460, entitled "Boston-Cambridge-Quincy, MA-NH," to the retrieval for "Boston, MA" as a separate variable. Figure 5 shows the effect of this addition to the same set as used for Figure 4. As can be expected, the larger region is more central than the city at the level of the global set. However, correlations between the two portfolios (Boston as a city and as CBSA) as distributions of patents over patent classes are large and highly significant: Pearson's $r = .984$ ($p<0.01$); Spearman's $\rho = .835$ ($p<0.01$); cosine = 0.984. The factor analysis using Boston CBSA instead of the city is virtually the same (Table 8; cf. Table 6). The number of patents in the CBSA is almost three times larger than that for the city itself (Table 2 above).

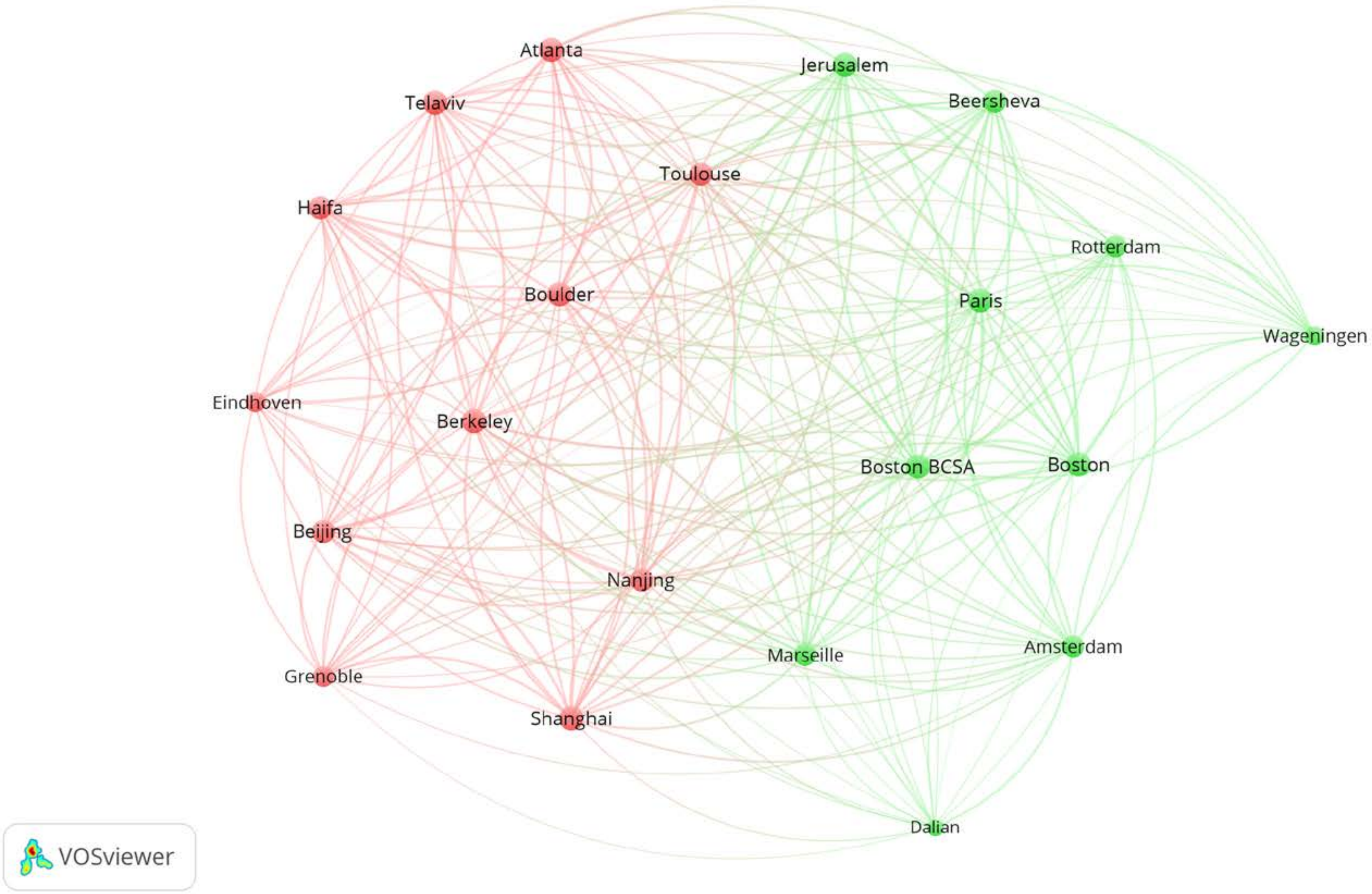

**Figure 5**: Cosine-normalized network among 20 cities and the CBSA of Boston ; VOSviewer is used for the clustering and mapping. The map can be web-started at
http://www.vosviewer.com/vosviewer.php?map=http://www.leydesdorff.net/software/patents/cos_map2.txt&network=http://www.leydesdorff.net/software/patents/cos_net2.txt&label_size_variation=0.3&scale=1.40&colored_lines&curved_lines&n_lines=1000&line_size_variation=0.55

**Table 8**: Varimax-rotated factor matrix of the patent portfolios of the same 19 cities and the "CBSA Boston-Cambridge-Quincy, MA-NH" instead of "Boston, MA" as a single city address.

**Rotated Component Matrix**[a]

| | Component | | |
|---|---|---|---|
| | 1 | 2 | 3 |
| Beijing | **.921** | .177 | .107 |
| Haifa | **.903** | .286 | |
| Berkeley | **.885** | .336 | |
| Tel-Aviv | **.877** | .374 | |
| Atlanta | **.860** | .362 | |
| Boulder | **.822** | .372 | |
| Shanghai | **.802** | .360 | .217 |
| Nanjing | **.737** | .369 | .273 |
| Grenoble | **.727** | | .132 |
| Toulouse | **.662** | .456 | |
| Eindhoven | **.612** | | |
| Marseille | **.525** | .495 | .171 |
| Wageningen | | **.823** | |
| Paris | .474 | **.807** | .152 |
| Rotterdam | .223 | **.802** | |
| CBSA Boston | .544 | **.776** | .170 |
| Beersheva | .455 | **.774** | |
| Jerusalem | .614 | **.727** | |
| Dalian | | | **.952** |
| Amsterdam | .243 | **.622** | **.636** |

Extraction Method: Principal Component Analysis.
Rotation Method: Varimax with Kaiser Normalization.
a. Rotation converged in 5 iterations.

Three factors explain 78.5% of the variance (vs. 78.3% in Table 5 above).

In summary, the structure of the matrix is not different; the vector for the city of Boston is structurally similar to that of the CBSA of Boston, whereas the *N* of granted patents is almost three times larger.

## 5. Discussion and Concluding Remarks

A number of recent studies have employed patent data, in particular patent classification codes, for the development of metropolitan and regional knowledge spaces (Kogler *et al.*, 2013; Rigby, 2015; Kogler *et al.*, 2016). The primary focus in these inquiries is on the evolution of regional knowledge spaces, while the analysis is based on measures derived from co-occurrence matrices of IPC codes. In the present study, a similar approach has been followed, but here the aim was to empirically specify the unique characteristics of technological portfolios of cities in terms of technological proximity and related variety.

First, a new instrument for the purpose of mapping and analysing patent portfolios of cities was introduced, and second, an analytical framework was suggested that allows for the statistical comparison of knowledge space properties amongst entities; in this case we resorted to cities as units of analysis. This further step is highly relevant from a policy perspective: the prospect of capturing, analysing, and comparing the technological knowledge competencies of a specific city vis-à-vis other places provides the opportunity for policy-makers and other stakeholders to identify the most promising avenues for deepening the local knowledge base as well as where to invest, what is usually limited resources, for further technological upgrading (Heimeriks and Balland 2015). The examples outlined above demonstrate that cities have very different and unique technological portfolios. Given this variation, a 'one size fits all' policy at the national level to further developing the technological knowledge base of cities can be counterproductive.

Although the results provide the opportunity for comparing relevant peer cities, this information needs to be supplemented with contextual information. This includes the cities' particular strategies and missions, qualitative information regarding the institutional similarities between the cities in question, the relative location, but also the relative position in the hierarchy of technological advancement. In this way, the suggested approach can be used as a tool to benchmark a city in comparison to relevant peers, which in turn may help to identify relevant best-practices in well-performing cities that are otherwise comparable in terms of their knowledge base and specialization patterns.

The results of the discriminant analysis indicated that national institutional settings are an aspect of understanding the patent portfolio of cities. Frequently urban centres belonging to the same country also display similar positions in the knowledge space. This is in line with previous theoretical and qualitative case study insights emanating from the national innovation systems literature (Lundvall, 1988; Nelson, 1993). However, these results also connect to the argument in the literature (Kogler, 2015) that developments are specific since path-dependent (Martin & Sunley, 2006). In summary, patent portfolios of cities can be expected to be both geographically tainted and historically specific. Our methods may be less appropriate for the specification of disruptive forms of technological renewal that may, among other things, lead to lock-in into patterns of technological decline.

In other words, the idea that cities within the same national jurisdiction, i.e. the same national system of innovation, are predominantly located in close vicinity in the knowledge space points to certain degree of national rather than just place-specific path-dependency. This in turn links back to the instance that countries frequently pursue a common national science- and technology-policy approach that is then generically applied in a top-down fashion to all localities within the territory, while in reality place-based specific policies would require a bottom-up approach that takes into consideration the knowledge competencies that already exist, which consecutively would allow for identifying the most promising future local development pathways.

From this perspective, measures of technological distance create an understanding of the adjacent possibilities for further knowledge production that is available for diversification (Kogler *et al.*, 2016; Boschma *et al.*, 2015). For medium-sized cities such as Toulouse, an obvious strategy seems to be to identify options for related diversification. Given its strong pattern of specialisation, adjacent technological opportunities can clearly be identified. For Paris, however, its advantage lies in the diversity of its technological knowledge base. In addition to expanding its many technological strengths through related diversification, Paris seems well positioned to further develop a comparative advantage in complex technological knowledge that requires a recombination of diverse technological building blocks at both the global and local levels.

## Appendix I: Portfolio Analysis and Maps in terms of Patent Classes

### 1. Preparing input files

a. Download the following files from http://www.leydesdorff.net/ipcmaps into a single folder:
   - ipc.exe;
   - ipc.dbf (with basis information about the classes);
   - uspto1.exe (needed for the downloading of USPTO patents);
   - cos_ipc3.dbf and cos_ipc4.dbf (needed for the computation of distances on the map);

b. Run ipc.exe.

### 2. Options within ipc.exe

a. The program asks for a short name (≤ 10 characters) in each run. This name will be used as the variable name in later parts of the routine;

b. The first option is to download the patents from USPTO at http://patft.uspto.gov/netahtml/PTO/search-adv.htm ; detailed instructions for the downloading can be found at http://www.leydesdorff.net/ipcmaps;

c. USPTO has a maximum of 1000 records at a time; but one is allowed to follow-up batches; after each download, save the files in another folder or as a zip file;

### 3. The incremental construction of the files matrix.dbf and rao.dbf

a. After each run, a column variable is added to the (local) file matrix.dbf containing the distribution of the 630 CPC/IPC classes in the document set under study. If the file matrix.dbf is absent, it is generated *de novo* and the current run is considered as generating the first variable; matrix.dbf can be read by Excel, SPSS, etc., for further (statistical) analysis;

b. Similarly, a row variable is added after each run to the file rao.dbf containing diversity measures (explained in the article) as variables. This file is also de novo generated if previously absent. Distances are based on [1 – cos($x$,$y$)] for each two distributions **x** and **y**;

c. The routine ipc2cos.exe reads the file matrix.dbf and produces cosine.net and coocc.dat as (normalized) co-occurrence matrices that can be used in network analysis and visualization programs such as Pajek or UCInet.

### 4. Output files in each run

a. Two input files (vos3.txt and vos4.txt) are generated for mapping the portfolio at the three- or four-digit level of CPC/IPC, respectively, using VOSviewer; the distances and colors (corresponding to clusters) in the maps are based on the base-map (Leydesdorff, Kushnir, & Rafols, 2014);

b. Two input files (ipc3.vec and ipc4.vec) can be used as vectors in Pajek files provided at http://www.leydesdorff.net/ipcmaps . This allows for layouts other than VOSviewer and for more detailed network analysis and statistics; the files ipc3.cls and ipc4.cls are so-called cluster files which can be used in Pajek, among other things, for the extraction of the local maps at the respective levels.

c. The various fields in the USPTO records are organized in a series of databases that can be related (e.g., in MS Access) using the field "nr".